\documentclass[10pt,amssymb,twocolumn]{revtex4}
\usepackage{hyperref}
\usepackage{amsmath}
\usepackage{graphicx}
\usepackage{amssymb} 
\usepackage{amsfonts} 
\usepackage{epsfig}
\usepackage{float}

\begin{document}


\title{Disproof of Joy Christian's ``Disproof of Bell's theorem''}

\author{Florin Moldoveanu}
\email{fmoldove@gmail.com}
\affiliation{Committee for Philosophy and the Sciences, University of Maryland, College Park, MD 20742}


\begin{abstract}
Four critical elementary mathematical mistakes in Joy Christian's counterexample to Bell's theorem are presented. Consequently, Joy Christian's hidden variable model cannot reproduce any quantum mechanics results and cannot be used as a counterexample to Bell's theorem. The mathematical investigation is followed by a short discussion about the possibility to construct other hidden variable theories. 

A tutorial section on relevant Clifford algebra topics was added at the end to help interested readers decide for themselves the validity of Joy Christian's claims. Also an appendix section discusses recent developments.
\end{abstract}


\maketitle

\subsection*{Introduction} 

In a series of papers \cite{JoyDisproof1, JoyReply1, JoyConsolidations1, JoyIllusion1, JoyBell1, JoyBound1, JoyDisproof2, JoyCausality1}, Joy Christian claims to have constructed a local and realistic hidden variable theory able to reproduce quantum mechanics' correlations in the case of an EPR-Bohm experiment with spin one half particles. Joy Christian's model was received with serious criticisms \cite{Marcin1}, \cite{Holman1}, \cite{Grangier1}, \cite{Tung1}, \cite{Moldoveanu1}, but no single analysis could provide a decisive argument against it. Interestingly, the first comment \cite{Marcin1} came very close to uncover the first mathematical problem discussed below. In his reply \cite{JoyReply1} Joy Christian stated: ``With hindsight, however, it would have been perhaps better had I not left out as an exercise an explicit derivation of the CHSH inequality in Ref.~[1]. Let me, therefore, try to rectify this pedagogical deficiency here.''. This reply unfortunately managed to discourage other people from trying to check the validity of the mathematical results which are actually very easy to find.

Here are some key statements from subsequent critics' replies showing that the mathematical correctness was not the main focus of the criticisms:

``the model formally manages to reproduce some quantum theoretical expectation values correctly'' \cite{Holman1} 

``and here we will assume that the content of these papers is correct'' \cite{Grangier1}

``The first comment on Christian's paper, [9] by Marcin Pawlowski, is also different from the current paper. That comment seems to say that in Clifford-algebra-valued hidden variable theory it is unable to derive Bell's inequalities. This is not true since they are indeed derivable, as is explicitly shown in [2].'' \cite{Tung1}

"By providing an explicit factorizable model, Joy Christian's example only disproves the importance of Bell's theorem as an argument against contextual hidden variable theories.'' \cite{Moldoveanu1}.

In the meantime, a new challenge to Joy Christian's work came from informal physics blogs: if Joy Christian's claim of a local realistic theory able to reproduce quantum correlations is right, then a computer simulation on a classical computer would be possible. The author of this paper proceed to do just that: model Joy Christian's theory on a computer to clarify its claims of local realism. In order to translate the model to computer code, all results had to be systematically double checked and this led to the big surprise of finding out the elementary but critical mathematical mistakes presented below. 

\subsection*{Error 1: losing part of the correlation result by incorrect averaging}

In his first paper \cite{JoyDisproof1}, Joy Christian introduces his hidden variable $\mu$ as a random handedness of the basis for his geometric algebra basis. If $\{e_1, e_2, e_3\}$ are a set of fixed orthonormal vectors, the hidden variable $\mu$ is defined as follows
\begin{equation}
\mu \triangleq \pm I = \pm e_1 e_2 e_3 = \pm e_1 \wedge e_2 \wedge e_3
\end{equation}
(see Eq.~14 of Ref.~\cite{JoyDisproof1})

Then Joy Christian proceeds on computing the Clifford product of two bivectors $\mu \cdot a$, $\mu \cdot b$ obtaining in Eq.~17:

\begin{equation}
(\mu \cdot a)(\mu \cdot b) = - a \cdot b - a \wedge b
\end{equation}
which is correct (please note that the plus and minus factors in the two $\mu$'s cancel each other out and the expression is equivalent with $(I \cdot a)(I \cdot b)$). However, the last line of Eq.~17 is incorrect. By a well-known identity (Hodge duality): 
\begin{equation}
a \wedge b = I \cdot (a \times b)
\end{equation}
it is now easy to see that  the last line of Eq.~17 is incorrect when $\mu = -I$. Because $I$ is incorrectly replaced by $\mu$ and gains an illegal minus sign when $\mu = -I$, this leads to an incorrect canceling of the $a \wedge b$ term when averaging over all $\mu$'s in the oriented vector manifold ${\cal{V}}_3$.

In Ref.~\cite{Marcin1}, Pawlowski criticized Christian's proposal for the presence of Clifford Algebra valued observables if we are to get a scalar in the RHS of the CHSH inequality. Joy Christian responded \cite{JoyReply1} by stating that the average on ${\cal{V}}_3$ removes the $a \wedge b$ element. However, Eq.~3 of Ref.~\cite{JoyReply1} is incorrect just like the last line of Eq.~17 in Ref.~\cite{JoyDisproof1}

Even without spelling in detail the error, it is easy to see that the exterior product term should not vanish on any handedness average because handedness is just a paper convention on how to consistently make computations. For example one can apply the same incorrect argument to complex numbers because there is the same freedom to choose the sign of $\sqrt{-1}$ based on the two dimensional coordinate handedness in this case. Then one can compute the average of let's say $z = 3+2i$ for a fair coin random distribution of handedness and arrive at the incorrect answer: $<z> = 3$ instead of $<z> = z$. 

One advertised strength of geometric algebra is the ability to make computations in a coordinate-free fashion. If breaking up an object in its components and performing an average results in elimination of some components, then we are guaranteed that the operation is mathematically illegal.

The same mistake is present in the minimalist paper \cite{JoyDisproof2} in Eq.~4. Given a {\it fixed} bivector basis $\{\beta_i, \beta_j, \beta_k\}$, Eq.~3 of Ref.~\cite{JoyDisproof2} is correct. The corresponding $\lambda$-dependent basis product however should be the same as Eq.~3 of Ref.~\cite{JoyDisproof2} because $\beta_i (\lambda) \beta_j (\lambda) = \beta_i \beta_j \lambda^2  = \beta_i \beta_j$ and not gain an illegal $\lambda$ term for the cross product.

So how it was possible to have such an elementary mistake undetected? Most of Joy Christian's papers suffer from a convention ambiguity: in some cases the computations are done using the $\mu = \pm I$ convention with $I$ arising from a {\it fixed} basis, while in others the computations are done using the $\mu = I$ convention (indefinite Hodge duality) which means that $I$ is the current trivector and does not arise from a fixed basis. Illegally mixing the conventions during one computation yields the supposedly agreement with quantum mechanics.

Another way this mistake can arise can be seen in Eqs.~23 and 24 of Ref.~\cite{JoyBound1}. In there Eq.~23 is correct and Eq.~24 is derived by switching $I$ to $-I$ for a change of handedness. It is true that changing handedness (or equivalent performing a reflection) changes the sign of the pseudoscalar $I$, but what is incorrect in Eq.~24 is that $a \times b$ is a pseudovector who should change sign as well. The corrected Eq.~24 should be:
\begin{equation}
(-I \cdot a)(-I \cdot b) = - a\cdot b - (-I)(-(a \times b))
\end{equation}

The same mechanism for producing the error probably occurred in deriving  Eq.~4 of Ref.~\cite{JoyDisproof2} due to an incorrect replacement of $\beta (\lambda)$ in Eq.~3 without appropriately switching the sign of the Levi-Civita pseudotensor.

Also there are two additional physics objections as well. First, associating a hidden variable to an abstract computation convention is completely unphysical. Second, by doing it so, the theory predicts the same correlation {\it regardless} of the spin state of the particles in the EPR-Bohm experiment.

\subsection*{Error 2: Isotropically weighted averages of non-scalar part of correlations and measurement outcomes cannot be both zero}

Specifically, Eq.~18 and Eq.~19 of paper \cite{JoyDisproof1} cannot be both right. Let us count how many factors of $\mu$ are in Eq.~18 and Eq.~19. In Eq.~18 there are two factors of $\mu$, one from $\mu \cdot n$ and the other from $d \rho (\mu)$. Eq.~19 has three factors of $\mu$ from $\mu \cdot a$, $\mu \cdot b$, and $d \rho (\mu)$. As the factors are even and odd, integrating on a manifold ${\cal{V}}_3$ where $\mu$ changes signs evenly, only one of the two equations can be zero. Expanding the $(\mu \cdot a)(\mu \cdot b)$ term and using Hodge duality it follows that isotropically weighted averages of non-scalar part of correlations and measurement outcomes cannot be both zero. 

By the prior error we already know Eq.~19 is incorrect and the statement that Eq.~18 and Eq.~19 cannot be both right is not a surprise. The new content of this error is that fixing Eq.~19 by any hypothetical generalization of the manifold ${\cal{V}}_3$ breaks Eq.~18. Both Eqs.~18 and 19 are needed to be right if the model is to reproduce experimental results. Therefore the handedness mistakes conclusively rule out both Joy's model and all its potential Clifford algebra generalizations.

\subsection*{Error 3: Illegal limit for a bivector equation}
In Ref.~\cite{JoyCausality1}, Joy Christian attempts to implicitly answer Holman's criticisms \cite{Holman1} that the final answer in Eq.~19 of Ref.~\cite{JoyDisproof1} has a wrong sign and that the outcome of the experiments is always the same resulting in perfect correlations. Joy Christian first seems to agree that the outcome for any pairs of experimental results is the same, but then tries to prove the opposite in an explanation marred by mathematical mistakes.

The discussion of this problem takes place around Eqs.~42-46 of that paper. Citing Joy Christian: {\it ``Furthermore, we have taken the randomness $\mu = + I$ or $- I$ shared by Alice and Bob to be the initial orientation (or handedness) of the entire physical space, or equivalently that of a 3-sphere. Consequently, once $\mu$ is given as an initial state, the polarizations along all directions chosen by Alice and Bob would have the same value, because $\mu$ completely fixes the sense of bivectors $\mu \cdot n$ belonging to $S2 \subset S3$ , regardless of direction.''} In other words, the correlation of the experimental outcome is always $+1$ contradicting quantum mechanics predictions. Up to this point Joy Christian is correct.

Still, Joy Christian continues: ``However, and this is an important point, the polarization $( +\mu \cdot a )$ observed by Alice is measured with respect to the analyzer $(- I \cdot a )$, whereas the polarization $( +\mu \cdot b )$ observed by Bob is measured with respect to the analyzer $(+ I \cdot b )$.'' 

It will be shown below in this and next section that the mathematical arguments supporting this (in an attempt to produce both the minus cosine correlation and all the four experimental outcomes $(+,+), (-,-), (+,-), (-,+)$) are mathematically incorrect.

First let us note that the change in sign between Alice and Bob is illegal because they both use the same kind of apparatus during measurement and swapping them should not change anything. Therefore we could stop the analysis here because Joy Christian's model is obviously wrong as it does not respect this basic symmetry. However let's follow along Joy Christian's argument and discover where the mistakes occur.

To solve the problem, Joy Christian considers two almost parallel vectors instead of one at each detector: $a a^{'}$ for Alice and $b b^{'}$ for Bob. The two vectors form a bivector and starting from an aligned vector configuration ($a$ aligned with $b$ and $a^{'}$ aligned with $b^{'}$) the goal is  to move the second pair ($b b^{'}$) at  Bob's location in the final detector position using a rotor. In general, a rotor can be expressed as follows:
\begin{equation}
R = \cos \Omega + \hat{B} \sin \Omega 
\end{equation}

with $\hat{B}$ a unit bivector defined as:
\begin{equation}
\hat{B} = \frac{m \wedge n}{\sin~ \Omega} 
\end{equation}
where $m$ and $n$ are two unit vectors and $\Omega$ is the angle between them. In the case of Eq.~45 of Ref.~\cite{JoyCausality1}, the unit bivector corresponds to an axial vector $c$ of unit norm:
\begin{equation}
c = \frac{a \times a^{'}}{|a \times a^{'}|} 
\end{equation}

Then at the end of the rotation, after applying a simple trigonometric identity composing bivectors (multiplying two exponential expressions corresponding to the $a \wedge a^{'}$ and the $a \wedge b$ bivectors) Joy Christian takes the limit $a \rightarrow a^{'}$ and claims that this makes the bivector component of the final result zero as $a \wedge a^{'}$ becomes zero. The end result is that only the cosine factor survives the operation therefore a rotation by ``parallel transport'' in a ``twisted manifold'' allows to recover the cosine of the angle between Alice and Bob in their correlation as predicted by quantum mechanics.

The limit operation above is mathematically illegal. To see why, recall that the axial vector $c$ is of unit norm and as such it is normalized by the sine of the angles between $a$ and $a^{'}$. As $a^{'}$ approaches $a$, the wedge product goes to zero as sine of the angle, but the denominator goes to zero by the same sine of the angle factor. As such the two sines cancel each other and the bivector maintains its magnitude. This is nothing but a restatement of the geometric algebra fact that the magnitude of a bivector does not depend on the shape of the parallelepiped defining it. But maybe there is a discontinuity at the limit when $a = a^{'}$ and the bivector $\hat{B}$ does become zero. We can see that this is not the case as follows. Suppose $a$, $a^{'}$, $b$, $b^{'}$ are four unit vectors in the same plane, and vector $c$ is a unit vector orthogonal to this plane:
\begin{equation}
c = \frac{a \times a^{'}}{|a \times a^{'}|}  = \frac{a^{'} \times b}{| a^{'} \times b|}
\end{equation}
let $\epsilon \Omega$ be the angle between $a$ and $a^{'}$, and $(1 - \epsilon) \Omega $ the angle between $a^{'}$ and $b$.
If $\hat{B}$ is the unit bivector
\begin{equation}
\hat{B} = B_{a a^{'}} = B_{a^{'}b} = I c = \frac{a \wedge a^{'}}{\sin (\epsilon \Omega)} = \frac{a^{'} \wedge b}{\sin (1 - \epsilon \Omega)} 
\end{equation}
then Joy Christian's incorrect argument is as follows:
\begin{eqnarray}
&\cos \Omega + \hat{B} \sin \Omega = \nonumber \\
&R(ab) = R(aa^{'}) R(a^{'}b) = \nonumber \\
&(\cos (\epsilon\Omega) + \hat{B}\sin (\epsilon\Omega))(\cos ((1- \epsilon)\Omega) + \hat{B}\sin ((1-\epsilon)\Omega)) = \nonumber \\
&\cos(\epsilon\Omega) \cos ((1- \epsilon)\Omega) - \sin(\epsilon\Omega) \sin ((1- \epsilon)\Omega) + \nonumber\\
&B_{a a^{'}}(\sin (\epsilon\Omega) \cos ((1- \epsilon)\Omega) + \cos (\epsilon\Omega) \sin ((1-\epsilon)\Omega)) = \nonumber \\
&\cos \Omega  \nonumber \\
\end{eqnarray}

The last equality comes from taking the exact limit $a = a^{'}$ which makes $B_{a a^{'}}$ vanish due to $a \wedge a = 0$. Comparing the first and last rows, it is clear that there is no discontinuity even when $a$ is strictly $a^{'}$ and the limit result is illegal.

It is also easy to see the mistake another way. Just compare Eq.~46 with the definition of the rotor following Eq.~45. In line two of Eq.~46 the term following the rotor computes to $-\lambda$ and the final answer in Eq.~46 up to the $-\lambda$ factor should be the entire value of the rotor and not just its cosine part.

\subsection*{Error 4: Incorrect parallel transport rotor direction}
In the problem above Joy Christian attempts to eliminate a bivector by an illegal limit. Taking the limit correctly still results in the wrong result because the first line of Eq.~46 should be equal with the last line of the same equation. The error is in using the incorrect rotor to perform the parallel transport. In geometric algebra any object $G$ transforms under a rotation by a rotor $R$ as $G \rightarrow R^{\dagger}GR$  with $R=ab$ and $R^{\dagger}=ba$. The angle of rotation is double the angle between vectors $a$ and $b$. This formula is completely general and works for scalars, vectors, bivectors, pseudoscalars, or any of their linear combinations. Let us try to apply this general formula to Eq.~46. $R_{ab}$ reads:
\begin{equation}
R_{ab} = \exp \{ (I \cdot c) \theta_{ab}/2\} = \cos (\theta_{ab}/2) + (I\cdot c)  \sin (\theta_{ab}/2)
\end{equation}

and the correct computation in Eq.~46 is:

\begin{eqnarray}
& \lim_{b^{'} \to b} [(+ I \cdot b) (+ \mu \cdot b^{'})] = - \lambda  \nonumber \\
& = \lim_{a^{'} \to a} \{R_{ab}^{\dagger} [(+ I \cdot a)  (+ \mu \cdot a^{'})] R_{ab}\} \nonumber \\
& = \lim_{a^{'} \to a}\{R_{ab}^{\dagger} [(-\lambda)(\exp \{ (I \cdot c) \theta_{aa^{'}}\})]  R_{ab}\} = \nonumber \\
& = \lim_{a^{'} \to a} \{R_{ab}^{\dagger} R_{ab} [(-\lambda)(\exp \{ (I \cdot c) \theta_{aa^{'}}\})]\} = \nonumber \\
& = \lim_{a^{'} \to a} [(+ I \cdot a) (+ \mu \cdot a^{'})] = -\lambda \nonumber \\
\end{eqnarray} 

The final result after parallel transport is still $-\lambda$ because $\lambda$ is a scalar. The fourth line in the equation above comes from the fact that vector $c$ commutes with itself. There is another way to understand why the final result was not changed even when the limit is not taken. A bivector is an oriented surface characterized only by direction, magnitude, and sense of rotation. The vectors $a$, $a^{'}$, $b$, $b^{'}$ are in the same plane and the bivector $[(+ I \cdot a) (+ \mu \cdot a^{'})]$ and $[(+ I \cdot b) (+ \mu \cdot b^{'})]$ are actually identical because they have the same orientation, magnitude, and sense of rotation.  Rotating the pair $a a^{'}$ into $b b^{'}$ by the angle $\theta_{ab}$ preserves the orientation, magnitude, and sense of rotation.

The second mistake in Eq.~46 (when computing the limit correctly) comes from applying incorrectly the law of rotation for rotors. This formula is different than the general multivector formula and it applies only for rotors. Specifically the mistake in Joy Christian's paper is in the orientation of his rotor $\cal{R}$: instead of rotating around the vector $c$ (or equivalently $a \times a^{'}$), the correct direction is $(a \times a^{'})\times ( b \times b^{'})$. This is the correct direction because we are trying to rotate an $a\wedge a^{'}$ bivector with orientation $a \times a^{'}$ into a $b\wedge b^{'}$ bivector with orientation $b \times b^{'}$ and the rotation needs to align the two directions. Therefore the correct rotation direction is $(a \times a^{'})\times ( b \times b^{'})$. With this correct direction one can prove that the rotor law of rotation and the multivector law of rotation produce the same result (for example one can use a tedious brute force expansion of the two formulae and show they are identical in components). Applied to the discussion in the paper, since $(a \times a^{'})$ is parallel with $(b \times b^{'})$ the rotor reduces to identity in this case.

Computed correctly with the right limit and the right rotor, we can now see that the outcome at Alice's and Bob's detectors is always the same and the results are completely correlated contradicting quantum mechanics.  The reason is that the outcome results are nothing but the negative of the local bivectors' magnitude - a fixed value regardless of direction. Holman's analysis \cite{Holman1} is therefore proven correct: ``Because $\mu$ is a local deterministic hidden variable, its value cannot depend on the choice made by the experimenter. If this value is left unchanged by the first measurement, performing the second measurement in the $e_x$-direction would result in ``spin up'' in this direction with certainty, in contradiction with the usual quantum predictions.''. Flipping of the signs on Alice's and Bob's experimental outcomes is also without merit.

\subsection*{Conclusion} 
Any of the four mathematical mistakes presented above can reject Joy Christian's claims of a disproval of Bell's theorem by counterexample. Another error is that computing Eq.~16 of Ref.~\cite{JoyBell1} yields Eq.~3 and not Eq.~15 as claimed: ``we believe the experiment will vindicate prediction (15) and refute prediction (3).''. This was proven both analytically and by computer simulation. Since this impacts only a particular claim of a paper and not the viability of the whole research program it was not presented here. (A similar computer simulation was carried out earlier by Stephen Lee \cite{Lee1}, but the source code was not made publicly available.) After those mistakes were found the systematic checking of Joy Christian's mathematical claims was stopped. 

Joy Christian's model is not correct, but can Bell's theorem be invalidated by another non-commutative ``beables'' theory? Two theorems by Clifton \cite{Clifton1} answer this in the negative for non-contextual hidden variable theories and for relativistic quantum field theories with bounded energy. During computer simulations, several other non-commutative models which correctly realize the minus cosine correlations were discovered. However, the simulation also showed that any transition from non-commutative beables to discrete experimental outcomes destroys this correlation and yields the classical correlations as expected from Bell's theorem. It is therefore essential for a hidden variable model to predict both quantum correlations and the experimental outcomes. Any commutative beable hidden variable theory is ruled out by Bell's theorem. Any non-commutative non-contextual beable hidden variable theory is ruled out by Clifton's analysis. The only remaining way out is to construct a non-commutative contextual beable hidden variable theory. But contextual hidden variable theories are not really considered physical theories as no experimental evidence ever backed them out. It is debatable if Bell's theorem is important to rule out some contextual hidden variable theories as well, or only non-contextual ones. Bell's theorem is not the only result ruling out non-contextual hidden variable theories, but only Bell's result is robust enough (because involves an inequality) to be put to an experimental test. 

\subsection*{Appendix: Relevant Clifford algebra tutorial}
After this preprint was originally published, several developments occurred. First, Joy Christian uploaded a rebuttal of this preprint \cite{JoyRebutalFlorin}, Richard Gill re-discovered error number one \cite{Gill1} in Joy Christian's minimalist one pager paper \cite{JoyDisproof2}, and Joy Christian rebutted that as well \cite{JoyRebutalGill}. There is no point going back and forth in an endless cycle of claims and counterclaims and instead this appendix will present in a hopefully clear pedagogical way the key ideas of Joy Christian's research, his mistakes (old and new), and why this program cannot be salvaged. The aim is to give any reader interested in this topic, even if not an expert in Clifford algebras, the tools to check Joy Christian's claims. 

Before starting I also want to make a minor correction based on Joy Christian's rebuttal. The only valid criticism in \cite{JoyRebutalFlorin} was that complex numbers don't have handedness. This is correct, and also completely inconsequential to the conclusions of this paper.

The hidden variable model in Joy Christian's proposal is based on Clifford algebra, also known as geometric algebra. In the $19^{th}$ century, Clifford himself called his algebra ``geometric algebra''. The term was popularized by David Hestenes in the 60s and Hestenes had a famous expression: ``Geometry without algebra is dumb! Algebra without geometry is blind! ''\cite{HestenesQuote}. But what is a Clifford algebra? One example is the algebra of Dirac's gamma matrices. Skipping the rigorous definition, a Clifford algebra is an algebra satisfying:
\begin{equation}
uv + vu = 2<u,v>
\end{equation} 
with the scalar product defined by the polarization identity:
\begin{equation}
<u,v>=(Q(u + v) - Q(u) - Q(v))/2
\end{equation} 
and $Q$ a quadratic form.

In particular Clifford algebra $Cl(3,0)$ used in Joy Christian's model satisfies:
\begin{equation}
e_i e_j + e_j e_i = 2 \delta_{ij}
\end{equation} 
with $(e_1, e_2, e_3)$ the usual $x, y, z$ unit vectors of the ordinary three dimensional space.

From this one extracts two very important practical rules:
\begin{equation}
e_1 e_1 = e_2 e_2 = e_3 e_3 = 1
\end{equation} 
and 
\begin{equation}
e_1 e_2 = -e_2 e_1 , e_1 e_3 = - e_3 e_1 , e_2 e_3 = -e_3 e_2
\end{equation} 

Therefore when computing expressions in $Cl(3,0)$, one has to keep proper track of the order of terms, and use the two practical rules to simplify the expressions. This is all there is to it and everything else follows from here. It is easy to see that the basis of $Cl(3,0)$ is: $\{1, e_1, e_2, e_3, e_1 e_2, e_2 e_3, e_3 e_1, e_1 e_2 e_3\}$ with $1$ being the scalar, $\{ e_1, e_2, e_3 \}$ the vector basis, $\{e_1 e_2, e_2 e_3, e_3 e_1 \}$ the bivector basis, and $e_1 e_2 e_3$ the trivector (pseudo-scalar). As a shorthand notation, the bivectors are represented by the exterior product $e_1 e_2 = e_1 \wedge e_2$, and the trivector is usually named $I = e_1 e_2 e_3$.

From the definition of Clifford algebra one can extract the scalar product:
\begin{equation}
a \cdot b = \frac{1}{2} (ab + ba)
\end{equation}
and from the definition of exterior product one has:
\begin{equation}
a \wedge b = \frac{1}{2} (ab - ba)
\end{equation}

Summing the two equations results in this fundamental identity:
\begin{equation}
ab = a\cdot b + a \wedge b
\end{equation}

In terms of geometry, the scalar corresponds to a point, the vectors to lines, the bivectors to oriented surfaces, and the trivector with an oriented parallelepiped. In general, in Clifford algebras Hodge duality maps $k$-vectors with $(n-k)$-vectors by multiplication with the pseudo-scalar. In particular, in $Cl(3,0)$, Hodge duality maps bivectors (oriented areas) to pseudo-vectors (the cross product vector orthogonal to the area):
\begin{equation}
a \wedge b = +I (a \times b)
\end{equation}

Now we have all the mathematical preliminaries to understand Joy Christian's model. Let us compute the following expression: $(I a)(I b)$ with $a$ and $b$ vectors. First, what is $I I$? Let us apply the two practical rules: $I I = e_1 e_2 e_3 e_1 e_2 e_3 = - e_1 e_2 e_3 e_1  e_3 e_2 = + e_1 e_2 e_3  e_3 e_1 e_2 = + e_1 e_2 e_1 e_2 = - e_1 e_1 e_2 e_2 = -1$. Then let us prove $Ia = aI$. We can show it component by component and we will compute only the ``x'' component leaving the rest as an exercise to the reader: $I a_1 e_1 = a_1 I e_1 = a_1 e_1 e_2 e_3 e_1 = + a_1 e_1 e_1 e_2 e_3 = a_1 e_1 I $ by hopping $e_1$ twice from right to left. So now $(I a)(I b) = I I a b = -ab = -a\cdot b -a\wedge b = -a\cdot b - I a \times b$. 

But why do we care about the computation above? Because the EPR-B singlet correlation is $-a\cdot b$ with $a$ and $b$ the spin measurement directions for Alice and Bob. If we can somehow eliminate the $-I a \times b$ term, we could claim we have a hidden variable theory reproducing quantum mechanics correlations. {\bf  This is the essence of Joy Christian's program. The essence of this tutorial is that this hope cannot be realized in any mathematical consistent way.} 

So let us follow Joy Christian again and introduce the hidden variable $\mu = \pm I$:
\begin{equation}
(Ia)(Ib) = -a\cdot b - \mu a\times b
\end{equation}
and the sign of $\mu$ being $+1$ $50\%$ of the time and $-1$ the other $50\%$ of the time. Is this equation correct? 

At first, Joy Christian's program seems feasible. The sign of the pseudoscalar/trivector $I$ is arbitrary and corresponds to the handedness of the reference frame $e_1, e_2, e_3$. So maybe the hidden variable is the sign of $I$ and we can average out somehow the troubled $I (a \times b)$ term. To see why this is impossible we need to clarify in the process the confusions between handedness, the sign of the determinant ($O(n)$ vs. $SO(n)$) and adjoin representations. 

During the many papers presenting his model, Joy Christian attempted several approaches to average out the $\mu a \times b$ term. This is what Richard Gill calls: ``moving the bump in the carpet''. We'll follow along the ``bump'' in historical fashion from Joy Christian's papers. However, before we embark on this endeavor, there is a very simple way to see that this term cannot vanish. We start with $(Ia)(Ib)$ which equals $-ab$ and no matter how this is decomposed and averaged in components, in the end is still $-ab$ which remains $-a\cdot b -a \wedge b$ and the $a \wedge b$ term cannot be eliminated in any consistent mathematical argument. 

Now we will introduce a few algebra preliminary statements without proofs. All Clifford algebras have even subalgebras. Handedness exists only in three dimensions (it is a property of the human hand: ``lefty loosy, righty tighty''), but it is generalized to any dimensions to a definite determinant sign which preserves orientation and distinguishes between $SO(n)$ and $O(n)$ by mirror reflections. All associative algebras have matrix representations (matrix multiplication is associative). A $n \times n$ matrix can multiply a column vector (ket) from the {\it left}, or a row vector (bra) from the {\it right}. For complex numbers and the even subalgebras of $Cl(3,0)$ those left/right algebra representations are adjoints (the usual transposed and complex conjugation, but without complex conjugacy for complex numbers because the representations are over real numbers). In Geometric Algebra, those adjoints correspond to reversing the order of the $e_i$ elements in products. 

Now let start illustrating those statements with complex numbers both in Clifford algebra and in matrix representations. Let us start with this Clifford algebra in two dimensions: $\{1, e_1, e_2, e_1 e_2\}$. It has a  right subalgebra: $\{1, e_1 e_2\}$ and a left subalgebra $\{1, -e_1 e_2\}$.

Now consider two elements $A$ and $B$ as follows: 
\begin{equation}
A = a_1 + a_2 e_1 e_2  
\end{equation}
\begin{equation}
B = b_1 + b_2 e_1 e_2  
\end{equation}
Let us multiply them:
$AB = (a_1 + a_2 e_1 e_2)( b_1 + b_2 e_1 e_2) = a_1 b_1 + a_2 b_2 e_1 e_2 e_1 e_2 + a_1 b_2 e_1 e_2 + a_2 b_1 e_1 e_2$ and using the two simple Clifford algebra rules one gets: $AB = (a_1 b_1 - a_2 b_2) + (a_1 b_2 + a_2 b_1) e_1 e_2$ so this is a Clifford algebra representation of complex numbers.

Similarly consider
$M$ and $N$ as follows: 
\begin{equation}
M = m1 +m2 (-e_1 e_2)  
\end{equation}
\begin{equation}
N= n1 + n_2 (-e_1 e2) 
\end{equation}
and after multiplication $MN = (m_1 n_1 - m_2 n_2) + (m_1 n_2 + m_2 n_1) (-e_1 e_2)$, so again this is the other a Clifford algebra representation of complex numbers. Please note that it is nonsense to add or multiply numbers in the two representations. Also the two representations are adjoint: ${(e_1 e_2)}^{\dagger} = e_2 e_1 = -e_1 e_2$.

Now let's illustrate complex numbers in the two left/right matrix and ket/bra representations which correspond to the two above subalgebra representations.

The ket representation is as follows:
\begin{equation}
| z > = \left( \begin{array}{c}
a\\
b\\
\end{array}
\right)
\end{equation} 
with this multiplication rule
\begin{equation}
| z_1 > * | z_2 > = \left( \begin{array}{c}
a_1 a_2 - b_1 b_2\\
a_1 b_2 + a_2 b_1\\
\end{array}
\right)
\end{equation} 
and the left matrix representation is
\begin{equation}
Z_{left} = \left( \begin{array}{cc}
a&-b\\
b&a\\
\end{array}
\right)
\end{equation} 

Then the following identities hold justifying the name (left representation)\cite{GrginBook}:
\begin{equation}
A B C | d > = A B |c d> = A B |c> * |d> = A |b c d>
\end{equation}

Similarly the bra representation is as follows:
\begin{equation}
< z | = \left( a , b \right)
\end{equation} 
with this multiplication rule
\begin{equation}
< z_1 | * < z_2 | = \left( a_1 a_2 - b_1 b_2 ,
a_1 b_2 + a_2 b_1 \right)
\end{equation} 
and the right matrix representation is
\begin{equation}
Z_{right} = \left( \begin{array}{cc}
a&b\\
-b&a\\
\end{array}
\right)
\end{equation} 

Then the following identities hold justifying the name (right representation):
\begin{equation}
< d | C B A = <d c | B A = <d| * <c| B A = <d c b | A 
\end{equation}

Please note:
\begin{equation}
|z>^{\dagger} = <z|
\end{equation}
and
\begin{equation}
Z_{left}^{\dagger} = Z_{right}
\end{equation}

The reason for this mathematical detour is to clarify the left/right mathematical representations because in the case of the even subalgebras of $Cl(3,0)$ Joy Christian incorrectly calls them left/right handedness \cite{JoyRebutalFlorin}. In general, for an associative algebra the left/right ket/bra representations are distinct from the adjoint representations, and also different from chiral representations. For complex numbers and for $Cl(3,0)$ subalgebras, the adjoint and left/right matrix representations are identical (as it was already shown for complex numbers and will be later shown for the even subalgebras of $Cl(3,0)$). In case of Joy Christian's subalgebras we have four distinct classes: left algebra left handedness, left algebra right handedness, right algebra left handedness, right algebra right handedness. Recall that a handedness change corresponds to a mirror reflection, and an adjoint change corresponds to flipping the order of the basis. So here are the four classes of algebras one can encounter in Joy Christian's model:
Start with a {\it fixed} right hand basis $\{ e_1 , e_2 , e_3\}$ 

Right algebra right handed basis:
\begin{equation}
\left\{ 1, e_2 e_3 , e_3 e_1 , e_1 e_2 \right\}
\end{equation}
then mirror say $e_2$ to get a right algebra in a left handed basis
\begin{equation}
\left\{ 1, (-e_2 ) e_3 , e_3 e_1 , e_1 (-e_2 ) \right\}
\end{equation}
From right/right flip the order of all basis vectors (take a geometric algebra adjoint) to get a left algebra in a right handed basis:

\begin{eqnarray}
\left\{ 1, {(e_2 e_3 )}^{\dagger} , {(e_3 e_1 )}^{\dagger}, {(e_1 e_2 )}^{\dagger}\right\} = \nonumber \\
\left\{ 1, e_3 e_2 , e_1 e_3 , e_2 e_1 \right\} = \\
\left\{ 1, - e_2 e_3 , -e_3 e_1 , -e_1 e_2 \right\} \nonumber 
\end{eqnarray}

Last, to obtain a left algebra in a left handed basis, flip the sign of $e_2$ for example in the equation above:
\begin{equation}
\left\{ 1, e_2  e_3 , (-e_3 e_1 ) , e_1 e_2 \right\}
\end{equation}

Let us now defines some useful shorthand notations: $B_{1R} = e_2 e_3$, $B_{2R} = e_3 e_1$, $B_{3R} = e_1 e_2$ and the corresponding adjoints $B_{L}=B_{R}^{\dagger}$: $B_{1L} = -e_2 e_3$, $B_{2L} = -e_3 e_1$, $B_{3L} = -e_1 e_2$. Let us compute the product of the three elements in both algebras: $B_{1R} B_{2R} B_{3R} = +1$ and $B_{1L} B_{2L} B_{3L} = -1$. For proof, please see this sequence $e_2 e_3 e_3 e_1 e_1 e_2 = e_2 e_3 e_3 e_2 = e_2  e_2 = +1$. Again, Joy Christian incorrectly calls this right and left handedness, but the name actually comes from the matrix representation, and mixing them in a computation amounts to adding column with row vectors.

The general rule for multiplying the $B$ elements is as follows:
\begin{equation}
B_{iR} B_{jR} = - \delta_{ij} - \epsilon_{ijk} B_{kR}
\end{equation}
and
\begin{equation}
B_{iL} B_{jL} = - \delta_{ij} + \epsilon_{ijk} B_{kL}
\end{equation}

The relations between the two algebras are:
\begin{equation}
B_{R}^{\dagger} = B_{L} = - B_{R}
\end{equation}

Below are the ket/bra and left and right matrix representations of those two algebras (over complex numbers). Other representations are possible. In particular the representations over real numbers involve $4 \times 4$ matrices. The way to obtain the complex representations are either from Pauli matrices (those algebras are actually quaternions and the $B$s are the $\pm i$, $\pm j$, $\pm k$ quaternionic elements) or from the Cayley-Dickson construction (which was used below). The advantage of the Cayley-Dickson construction is that it also gives the ket/bra representations right away.

Left algebra ($B_{iL} B_{jL} = - \delta_{ij} + \epsilon_{ijk} B_{kL}$):

\begin{equation}
| B_{1L} > = \left( \begin{array}{c}
i\\
0\\
\end{array}
\right)
\end{equation} 

\begin{equation}
| B_{2L} > = \left( \begin{array}{c}
0\\
1\\
\end{array}
\right)
\end{equation} 

\begin{equation}
| B_{3L} > = \left( \begin{array}{c}
0\\
-i\\
\end{array}
\right)
\end{equation} 

\begin{equation}
B_{1L} = \left( \begin{array}{cc}
i&0\\
0&-i\\
\end{array}
\right)  = i \sigma_3
\end{equation} 

\begin{equation}
B_{2L} = \left( \begin{array}{cc}
0&-1\\
1&0\\
\end{array}
\right)  = - i \sigma_2
\end{equation} 

\begin{equation}
B_{3L} = \left( \begin{array}{cc}
0&-i\\
-i&0\\
\end{array}
\right) = -i \sigma_1
\end{equation} 

Right algebra ($B_{iR} B_{jR} = - \delta_{ij} - \epsilon_{ijk} B_{kR}$):

\begin{equation}
< B_{1R} | = \left( -i , 0 \right)
\end{equation} 

\begin{equation}
< B_{2R} | = \left( 0 , 1 \right)
\end{equation} 

\begin{equation}
< B_{3R} | = \left( 0 , i \right)
\end{equation} 

\begin{equation}
B_{1R} = \left( \begin{array}{cc}
-i&0\\
0&i\\
\end{array}
\right) = -i \sigma_3
\end{equation} 

\begin{equation}
B_{2R} = \left( \begin{array}{cc}
0&1\\
-1&0\\
\end{array}
\right) = i \sigma_2
\end{equation} 

\begin{equation}
B_{3R} = \left( \begin{array}{cc}
0&i\\
i&0\\
\end{array}
\right) = i \sigma_1
\end{equation} 

By inspection, in ket/bra or left/right matrix representation, the same relationship holds as in the case of Clifford algebra representation: 
\begin{equation}
B_{R}^{\dagger} = B_{L} = - B_{R}
\end{equation}
and this time the adjoint involves also a complex conjugation as well because the representation is over complex numbers.

We are almost ready to discuss Joy Christian's mathematical mistakes and conclude that his program cannot succeed. Having clarified the left/right name and the relationship between Clifford $Cl(3,0)$ subalgebras in three representations, we need to understand how handedness enters the picture and how the sign of the Hodge duality changes or not in the four combinations: left/right representation with left/right handedness. To avoid unnecessary confusions we have considered and continue to consider only a {\it fixed} right handed basis $\{ e_1 , e_2 , e_3\}$. Any mirror reflection changes handedness and there is no restriction on which direction the mirror reflection can take place. We have seen that the left and right subalgebra representations obey different mathematical identities: $B_{1R} B_{2R} B_{3R} = +1$ and $B_{1L} B_{2L} B_{3L} = -1$. Let us first show that those identities have nothing to do with handedness contrary to Joy Christian's claims. Start with a right subalgebra in a fixed right handed basis $\{ e_1 , e_2 , e_3\}$. Perform three mirror reflections for $e_1$, $e_2$, and $e_3$. All three $e_i$ elements change signs and the $B$ elements do not because they are an even product of the unit vectors. As such the triple product $B_{1R} B_{2R} B_{3R}$ maintains the same $+1$ sign. But the handedness flipped three times in the process. Therefore the sign of the triple product has absolutely no relevance on the basis handedness (the sign of $I$).  Drawing a right hand basis on paper and successfully drawing the three consecutive mirror reflections shows how the final basis can change signs for the pseudoscalar.

Now the final piece of the puzzle: the sign in the Hodge duality. By direct computation in components and term identification, it is easy to see that in a right handed right algebra:
\begin{equation}
a \wedge b = +I (a \times b)
\end{equation}
The question is: does this equation changes signs by a mirror reflection as Joy Christian seems to imply in his first paper? Now we can either compute it in components, or observe instead that in a mirror reflection $I$ changes signs as a pseudo-scalar, but so does $a \times b$ because it is a pseudo-vector. As such, Hodge duality maintains its sign on a handedness change as long as the algebra is kept the same. Leaving the proof as an exercise to the reader, in a right handed left algebra the Hodge duality becomes:
\begin{equation}
a \wedge b = -I (a \times b)
\end{equation}
and again, this relation does not change signs on any mirror reflection.

So now we have all the mathematical results needed to understand the question posed earlier: is the following equation correct?
\begin{equation}
(Ia)(Ib) = -a\cdot b - \mu a\times b
\end{equation}

Now let us look in chronological order to Joy Christian's papers and see the "bump in the carpet" moving. Starting with the first paper \cite{JoyDisproof1}, we have a case of the same algebra and different handedness. But Hodge duality does not change signs on a handedness change in the same algebra and the last line of Eq.~17 of the paper is inconsistent with Eq.~14 of the same paper.

When confronted in public blog debates with the proof of this mathematical fact, Joy Christian used the ambiguity in his framework and claimed that this is a ``straw man'' argument and he actually meant left and right algebras. This is consistent with his rebuttal argument for this paper. Citing Joy Christian \cite{JoyRebutalFlorin}:

``To this end, right-multiply both sides of Eq. (7) by $\beta_l$, and then use the fact that ${(\beta_l)}^{2} = -1$ to arrive at
\begin{equation}
\beta_j \beta_k \beta_l = +1 \hspace{0.5in} [12] \nonumber
\end{equation}
The fact that this ordered product yields a positive value confirms that $\{\beta_x , \beta_y , \beta_z \}$ indeed forms a right-handed
frame of basis bivectors. This is a universally accepted convention, found in any textbook on geometric algebra [14].''

I hope now it is easy to spot the serious confusions in Joy Christian's text snippet. The overall sign of the triple product has nothing to do whatsoever with handedness. Instead it has to do with row or column representations or with multiplying kets from the left and bras from the right in matrix representations, and any handedness can be picked by whatever convention one wants. 

So if we no longer consider the handedness to change the sign of the Hodge duality, we can now investigate the next move in "bump in the carpet" as displayed in Joy Christian's one-pager paper \cite{JoyDisproof2} which Richard Gill investigated in \cite{Gill1}. In this paper, following Eq.~3, Joy introduces lambda-dependent bivector basis: $\beta_j ( \lambda ) = \lambda \beta_j$. Apart from this dangerously confusing notation, we can see that this is nothing but the relationship between the left and right algebras (ignore the left most term):
\begin{equation}
B_{R}^{\dagger} = B_{L} = - B_{R}
\end{equation}

At this juncture, it is hard to understand Joy Christian's intent as we reached a decision point: is Joy Christian mixing the two representations in the same computation, or not? Either way it leads to bad outcomes. In public blog debates, Joy Christian opted for the first choice: adding the two betas. If this is the case the entire model evaporates. By substituting $\beta_j ( \lambda ) = \lambda \beta_j$ into his Eq.~3, Joy Christian obtains his Eq.~4:
\begin{equation}
\beta_j \beta_k = -\delta_{jk} - \lambda \epsilon_{jkl} \beta_l 
\end{equation}

Now if this equation is always valid (for both values of lambda) by adding the two equations for both values of lambda one can immediately see that all betas must be zero. In this appendix formalism, when using the ket/bra representation, it is obvious that it is nonsensical to add row with column vectors and Joy Christian's Eq.~4 cannot be considered holding at the same time. 

So now let us move the "bump" along one more time and argue that in $50\%$ of the cases the equation with $\lambda = +1$ holds, and in the other $50\%$  of the cases the equation with $\lambda = -1$ holds.

Then the problem moves to deriving Eq.~7 from Eq.~6 in Joy Christian's one pager. Here the error is subtle and if the reader loves a mathematical challenge, please try to identify the issue without reading on.

To get from Eq.~6 to Eq.~7, Joy Christian is substituting Eq.~4 into Eq.~6. But while in Eq.~4 lambda is a variable, in Eq.~6 is an index and the substitution in one step is illegal. To do the proper substitution, we need to keep proper tabs on the two kinds of betas in Eq.~6: row or column vectors and add like terms in two substitutions. Then use the conversion law: $B_R = -B_L$ to combine the two sums. This additional minus sign from the conversion now makes the trouble sum term in Eq.~7 no longer vanish and the correct final result is (no surprise):  $-a \cdot b - a \wedge b = -ab$

So no matter which road one takes computing $-ab$, the final result must be $-ab$ and not $-a\cdot b$.

We have reached the end of the mathematical road hopefully illuminating the mathematics involved in Joy Christian's research program and despite its unusual Clifford algebra setting, the math involved is really trivial. It is a matter of indifference on which representation one chooses for $Cl(3,0)$ and its subalgebras, be it standard geometric algebra, ket and bra representations, or matrix representations. The practical rules given by Eqs.~16 and 17 are all one needs to know to be able to check Joy Christian's claims. A larger detour was given here to put things in perspective and understand the left and right representations, mirror reflection impact, and spelling out the four classes of algebras: left/right representation times left/right handedness.

\subsection*{Appendix: Final developments}
After writing the appendix above and before this paper was made public, important developments occurred changing completely the author's perspective. Tung Ten Yong's original analysis \cite{Tung1} invalidated the entire program of challenging Bell's by ``topological complete'' hidden variable models, but his argument was not understood until James Owen Weatherall managed to construct a model (not based on Clifford algebras) which achieves precisely the aims of Joy's program: $-a\cdot b$ correlation and zero average for both Alice and Bob's outcomes \cite{Weatherall1}. In particular, Weatherall's model respects outcome and parameter independence but it cannot be modeled on a computer according to Sascha Vongehr's quantum Randi challenge \cite{Vongehr1}. So what goes wrong as James Weatherall discovered independently of Tung Ten Yong, is the need to compute the correlations not in a generalized statistical theory in the space of the hidden variable model, but in the usual way {\it after} the experimental outcomes are obtained. The conclusion of those developments is that Bell's locality: outcome and parameter independence has to be augmented with the requirement to compute the correlations after obtaining the experimental outcomes. Therefore the Clifton's theorems \cite{Clifton1} are no longer needed to prevent challenges to Bell's theorem.
 
More developments followed. ``Error 1'' presented in this paper was discovered independently not only by Richard Gill and the author, but by Marc Holman and David Hestenes as well but they were not published. I did not want to change anything in the original sections of this paper to let the readers have a clear history of this debate. However, the earlier statement: `` but no single analysis could provide a decisive argument against it'' is clearly false and reflected the author's understanding at the time of writing the first version. All four original critics: Pawlowski, Holman, Grangier, and Yong provided decisive arguments (and Holman's analysis provided two independent decisive arguments against the model). Marc Holman even prepared a paper on ``Error 1'' \cite{MarcPrivate} which was never published. To date, Joy Christian still disagrees with all his critics, calling their arguments: ``straw men arguments'', and stating that he will not accept anyone's judgment on his work ``but that of Nature'' reiterating his call for the experiment proposed in \cite{JoyBell1}. In the author's opinion, demanding an experiment while failing basic mathematical consistency amounts to an unacceptable departure from basic scientific norms. Hopefully the appendix will allow the reader reach an informed decision on the mathematical consistency of Joy Christian's mathematical argument.   

\subsection*{Acknowledgement}
I want to thank Cristi Stoica for his help in computer modeling of Joy Christian's proposed experiment in Ref.~\cite{JoyBell1} and for useful discussions. I also want to thank Marc Holman, Richard Gill, David Hestenes, James Weatherall, and Sascha Vongehr for useful discussions. I want to thank Scott Aaronson for helping decide on the usefulness of writing the Clifford algebra tutorial.

\end{document}